\documentclass[sigconf]{acmart}

\AtBeginDocument{%
  }

\setcopyright{acmlicensed}
\copyrightyear{2024}
\acmYear{2024}
\acmDOI{}

\acmConference[MLCAD '24]{MLCAD}{Sept. 9-11,
  2024}{Snowbird, UT}

\acmISBN{979-8-4007-0699-8/24/09}

\usepackage{comment}
\usepackage{amsmath,amsfonts}
\usepackage{algorithmic}
\usepackage{graphicx}
\usepackage{textcomp}
\usepackage{multirow}
\DeclareMathOperator*{\argmax}{arg\,max}
\usepackage[section]{placeins}
\usepackage{dblfloatfix}
\usepackage{subcaption}
\usepackage{listings}
\usepackage{color}
\usepackage{tikz}
\usepackage[pages=some]{background}
\usepackage{booktabs}
\usepackage{longtable}
\usepackage{hhline}
\usepackage{colortbl}
\usepackage{tabularray}
\usepackage{tcolorbox}
\definecolor{blue}{rgb}{0.13, 0.13, 1}
\definecolor{red}{rgb}{0.9, 0.4, 0}
\definecolor{green}{rgb}{0, 0.5, 0}
\definecolor{gray}{rgb}{0.5, 0.5, 0.5}
\definecolor{black}{rgb}{0,0,0}
\definecolor{fog}{rgb}{0.6,0.84,1.0}
\definecolor{spring}{rgb}{0.61,0.8,0.61}

\begin{document}

\title[Rome was Not Built in a Single Step: Hierarchical Prompting for LLM-based  Chip Design]{Rome was Not Built in a Single Step: \\ Hierarchical Prompting for LLM-based  Chip Design}

\author{Andre Nakkab}
\affiliation{\institution{New York University}
\city{}
\state{}
\country{}}
\email{andre.nakkab@nyu.edu}
\orcid{0009-0006-1345-5444}

\author{Sai Qian Zhang}
\affiliation{\institution{New York University}
\city{}
\state{}
\country{}}
\email{sai.zhang@nyu.edu}

\author{Ramesh Karri}
\affiliation{\institution{New York University}
\city{}
\state{}
\country{}}
\email{rkarri@nyu.edu}

\author{Siddharth Garg}
\affiliation{\institution{New York University}
\city{}
\state{}
\country{}}
\email{siddharth.garg@nyu.edu}

\renewcommand{\shortauthors}{Nakkab et al.}

\begin{abstract}
  Large Language Models (LLMs) are effective in computer hardware synthesis via hardware description language (HDL) generation. However, LLM-assisted approaches for HDL generation struggle when handling complex tasks. 
We introduce a suite of hierarchical prompting techniques which facilitate efficient stepwise design methods, and develop a generalizable automation pipeline for the process. To evaluate these techniques, we present a benchmark set of hardware designs which have solutions with or without architectural hierarchy. Using these benchmarks, we compare various open-source and proprietary LLMs, including our own fine-tuned Code Llama-Verilog model. Our hierarchical methods automatically produce successful designs for complex hardware modules that standard flat prompting methods cannot achieve, allowing smaller open-source LLMs to compete with large proprietary models. 
Hierarchical prompting reduces HDL generation time and yields savings on LLM costs. Our experiments detail which LLMs are capable of which applications, and how to apply hierarchical methods in various modes. We explore case studies of generating complex cores using automatic scripted hierarchical prompts, including the first-ever LLM-designed processor with no human feedback.
\end{abstract}


\keywords{LLM, Hardware design, Hierarchy, Automation}


\maketitle
\backgroundsetup{opacity=1, scale=1, angle=0, contents={
\begin{tikzpicture}[remember picture, overlay]
\node[anchor=north east, inner xsep=50pt, inner ysep=10pt] at (current page.north east) {
\href{https://www.acm.org/publications/policies/artifact-review-and-badging-current}{
\includegraphics[width=50pt]{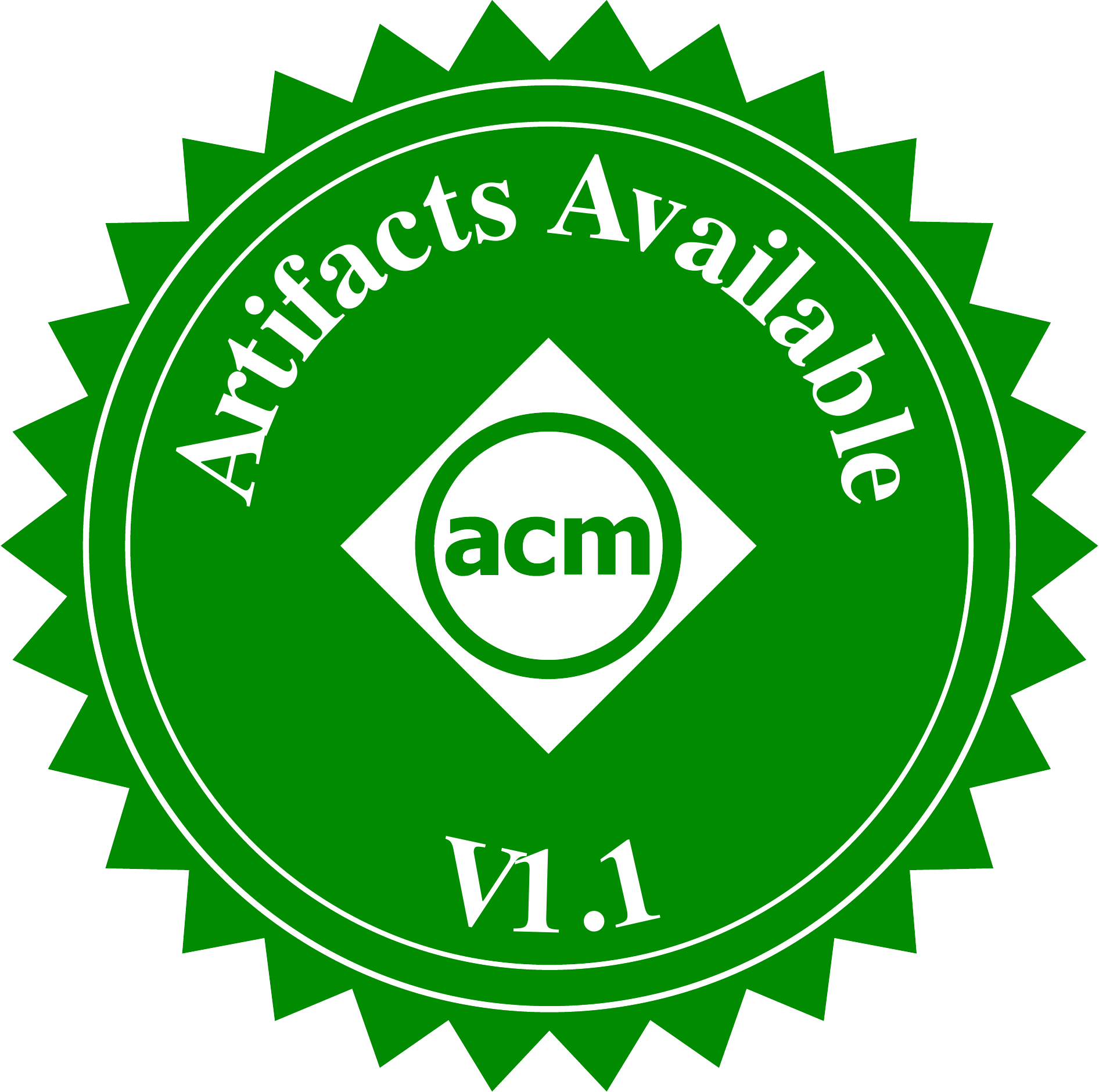}
}};
\end{tikzpicture}
}}
\BgThispage
\section{Introduction}
Hierarchical design is a key concept for creating complex computer hardware in an organized fashion. The goal of hierarchy is to break complex modules into manageable submodules, the way one might define a function in high-level code. 
However, recent efforts into LLM-based generation of hardware description language (HDL) code ~\cite{thakur2023verigen,liu2024rtlcoder,thakur2023autochip} generate modules non-hierarchically, i.e.,  as single blocks of straight-line code. 
Although these methods succeed on simple designs like bit-parallel adders and shift registers ~\cite{thakur2023benchmarking}, they 
struggle on complex designs in recent benchmarks, such as finite-state machines (FSM), large-scale many-to-1 multiplexers, and larger arithmetic blocks ~\cite{liu2023verilogeval}. Since straight-line code blocks for complex designs are longer than the hierarchical alternatives, 
they may hallucinate~\cite{liu2024exploring}; the LLM generates incorrect or unrelated text. Additionally, long outputs increase response latency and sometimes fail due to output length limits.

In this paper, we develop and evaluate \emph{hierarchical prompting} techniques to facilitate automated generation of \emph{modular} HDL code.
We explore hierarchical Verilog generation in two major modalities, each occurring in the real-world.
In the \emph{human-driven} mode, the prompt contains a human-proposed hierarchy that the LLM must extract and implement, as well as iterative compiler feedback from unit tests for each submodule. In the more challenging \emph{purely generative} mode, the LLM gets only a basic (non-hierarchical) prompt and therefore must make its own design decisions to implement the target module. 
We implement an 8-stage pipeline to automate these techniques in a generalizable fashion, which we refer to as Recurrent Optimization via Machine Editing (ROME). This allows an LLM to closely emulate human HDL development practices.

   \begin{figure}
        \centering
        \includegraphics[width=\columnwidth]{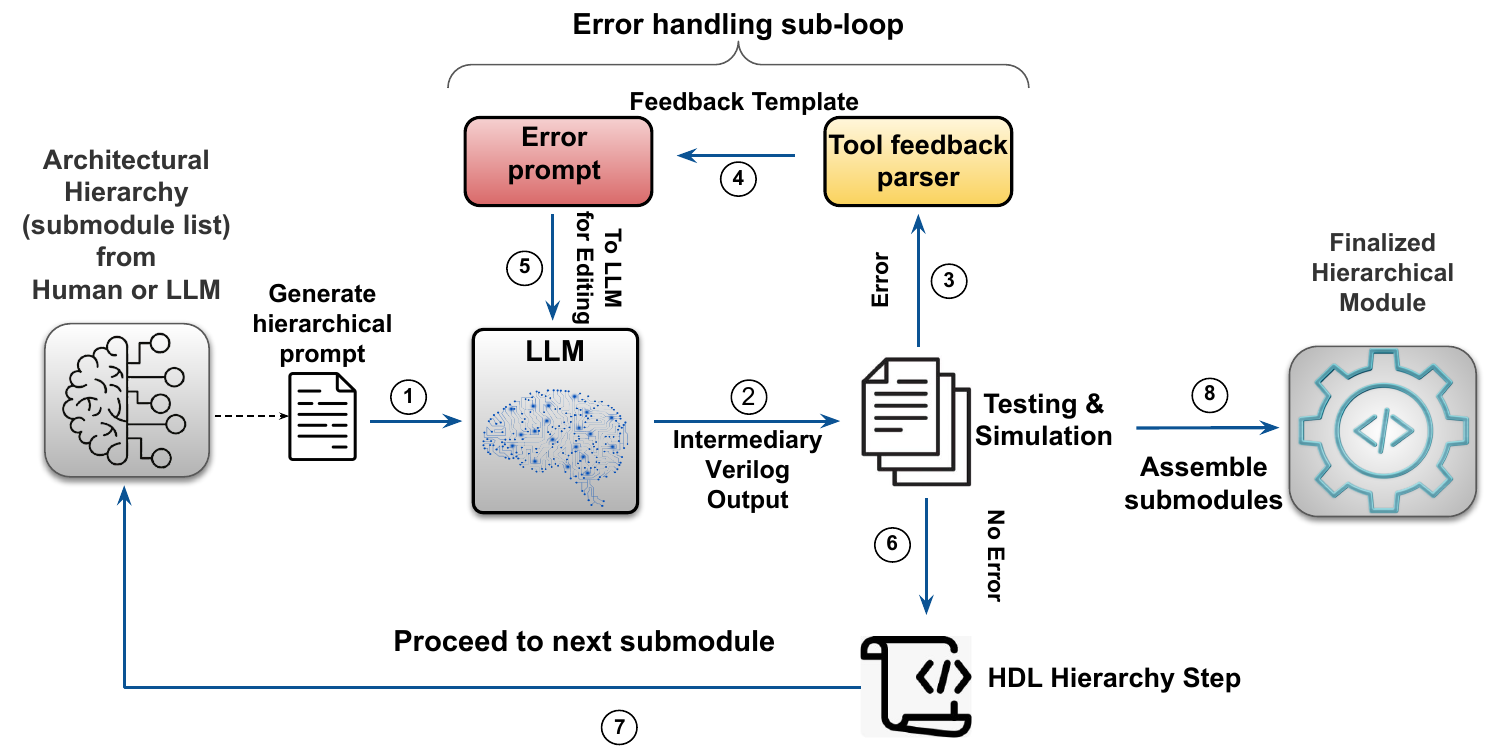}
        \caption{Automatic Hierarchical Prompting Pipeline.}
        \label{fig:flowchart}
        \vspace{-5mm}
    \end{figure}

Existing benchmark suites like VerilogEval~\cite{liu2023verilogeval} and RTLLM ~\cite{lu2023rtllm} do not address hierarchy. We introduce {\it a new benchmark suite of complex modules with explicit hierarchical solutions, including associated prompts and testbenches for both the top-level modules, and unit tests for submodules. } The target modules in our benchmark pose unique challenges to LLMs. These include a 32-bit data-dependent left rotation (also known as a barrel shifter) that requires rarely-seen syntax which is difficult to generate from scratch; an Advanced Encryption Standard (AES) block cipher which has multiple complex submodules that are difficult to organize within a single large Verilog script; and, a universal asynchronous receiver and transmitter (UART) interface which requires a multi-part FSM to function. These designs could not be implemented at all by any of the open-source or commercial LLMs we tried in non-hierarchical mode, motivating the need for advanced hierarchical prompting methods. 
The full list of benchmarks are in Table \ref{tab:bench_mods} in the Appendix.


 Evaluations on 8 state-of-art LLMs demonstrate that hierarchical prompt structuring dramatically improves LLM performance on hardware design tasks, enabling successful generation of modules that would otherwise be impossible. 
 Finally, we report case studies on the generation of complex hardware modules outside of the context of the hierarchical prompting benchmarks. We target a 16-bit MIPS processor and a 32-bit RISC-V processor, and present the first-ever purely LLM-designed processor with no human feedback.
 

\section{Background and Related Work}
\subsection{An Introduction of LLM Operation for Text Generation}
Transformer-based deep neural networks (DNN) have enabled advances across a wide range of domains, excelling in language-related tasks. LLMs operate by processing text inputs structured as \textit{tokens}, 
When presented with a sequence of input tokens, LLMs output a probability distribution spanning the complete \textit{vocabulary} to predict next token in the sequence. This process repeats until a full sequence of tokens, referred to as a \textit{completion}, is produced. Consider an LLM $P_{\phi}(\vec{y} | \vec{x})$, where $\phi$ denotes the set of model parameters, $\vec{x}$ and $\vec{y}$ represent the vector of input and output tokens. The LLM will generate the probability distribution for the next token $y_{n}$, and output $\vec{y} = \{y_{n}| 1\leq n \leq N\}$ is produced autoregressively:
\begin{equation}
    y_{n} = \argmax_{v\in V} P_{\phi}(v| \vec{x}, y_{<n})
    \label{eqn:autoaggressive}
\end{equation}
$1 \leq n \leq N$, $N$ is the length of the output, and $V$ is the set of vocabulary. Equation~\ref{eqn:autoaggressive} continues iterating until a designated end-of-sequence token is encountered.

\subsection{Related Work}
In recent years, LLMs have demonstrated their proficiency in code generation for software programming languages like C and Python~\cite{chen2021evaluating, nijkamp2022codegen, feng2020codebert, fried2022incoder, wang2021codet5,manyika2023overview,roziere2023code}.
This is possible because LLMs are trained using extensive datasets of code that encompass either one  programming language or a combination of multiple languages. Training datasets used can be substantial in size, reaching hundreds of gigabytes of text. Inputs supplied to these LLMs come in various formats, including instructions, comments, code excerpts, or some combination thereof. 
LLMs can further be tasked to generate hierarchical high-level models
~\cite{jiang2023self}. These approaches use Chain-of-Thought prompting to improve LLM reasoning by granularizing problems into sub-problems. This technique improves performance on mathematical word problems ~\cite{wei2023chainofthought}. Thus, LLMs are capable of understanding the functional intent of the design task at hand when the task is hierarchically structured. 
 
It is possible to use an LLM to conversationally generate synthesizable HDL at the processor-scale using feedback from a human hardware designer ~\cite{blocklove2023chip}.  These methods are relatively expedient, but struggle with consistency and are difficult to evaluate due to the subjective and non-reproducible nature of human feedback. Human intervention also precludes automation. It has alternatively been shown once in the past that one can automate the generation of entire CPUs using traditional deep learning methods ~\cite{cheng2023pushing}. However, these methods take on the order of multiple hours to successfully train and produce a given RTL design for tape-out. 
Finally, some success has been found by fine-tuning open-source LLMs specifically to produce HDL ~\cite{thakur2023verigen,liu2024rtlcoder}. Benchmarks have recently been developed to evaluate these LLMs on their Verilog generation performance ~\cite{lu2023rtllm,liu2023verilogeval}. Based on these benchmarks, even  bespoke fine-tuned models struggle to compete with powerful proprietary LLMs like GPT-3.5 and GPT-4. 

\section{Hierarchical Prompting}
Standard flat prompting involves straightforwardly asking the LLM to generate your desired module. This is effective for small, simple modules, but will often lead to messy, incorrect outputs when applied to more complex hardware structures. 
The goal of introducing architectural hierarchy via prompting is to allow an LLM to mimic the design process employed by a human engineer. Rather than writing every step of a given hardware element sequentially, we break it down into functional components and pick out reusable blocks that are simpler to produce, and slot them into the greater design.

\subsection{Sources of Hierarchy}
Hierarchical prompting can take a few forms depending on the resources available to the user. As a baseline, we can utilize a human-defined hierarchy as an input to the pipeline, which results in solely LLM-generated Verilog and no human involvement beyond the planning stage. This is effective when there are specific design constraints for the final module that the LLM might not recognize on its own. We refer to this method as \emph{human-driven hierarchical prompting} (HDHP).

At its most automated, hierarchical prompting can be used to generate our final module purely from the outputs of the LLM. We can describe some desired module and ask the LLM to give us a breakdown of the necessary blocks. From there, we can ask the LLM to generate the next block in the sequence. We refer to this method as \emph{purely generative hierarchical prompting} (PGHP). This is effective for the more common hierarchical modules, as information about them likely appears in standard LLM training datasets, but is a very difficult task when applied to rarely implemented modules. 

Our benchmark presumes HDHP-based design methods by default, as knowledge of the hierarchy allows us to create effective unit tests for each submodule in order to fully implement our feedback loop. Conversely, when utilizing PGHP the submodules which the LLM selects are consistent across runs. We can work around this by including a thorough, human-written testbench for the top module whose inputs \& outputs we know, and then optionally allowing the LLM to generate its own unit tests as we go.

\subsection{ROME Hierarchical Generation Pipeline}
As seen in Figure \ref{fig:flowchart}, we formulate an automatable design pipeline which implements hierarchical HDL generation to create a complete hierarchical hardware module from end to end. 
Our proposed method broadly works in three phases:
\paragraph{Hierarchy Extraction} In {Step 1}, we extract a list of submodules necessary to implement the design from a natural language description provided by the user.  Extraction  of submodules could be from a list provided by the user in HDHP mode, or extracted from the LLM in the more challenging PGHP mode.

\paragraph{Submodule Implementation} 
Next, we iterate through the extracted list of submodules, and in each iteration, ask the LLM to produce HDL for one
submodule (\emph{Step 2} in Figure~\ref{fig:flowchart}).
If unit tests for submodules are provided, the generated HDL is simulated via an HDL simulator, in our case Icarus Verilog (iVerilog) as it is open-source and easy to automate. Errors from the simulation output are then extracted and fed back to the LLM with an automated request to fix the design, forming a feedback loop that iteratively corrects errors in a given submodule.
(\emph{Steps 3, 4, 5} in Figure~\ref{fig:flowchart}).
Once a submodule passes tests, the generated code is logged, 
and we proceed to generate the next submodule (Steps 6,7). If no unit tests are provided as in the case of PGHP mode, the first generated submodule instance is picked.

\paragraph{Top-Level Module Integration} 
Finally, we request the LLM to integrate all generated submodules into a top-level module, which in turn has its own testbench. It is then put through the same tool-based feedback loop as the submodules, before finally being output as a completed hierarchical design. This is even possible in PGHP mode, as we know the expected behavior of the top-level module. We consider the run a success if the top module passes its testbenches.

\subsection{Prompting Structure and Techniques}
Within the pipeline, we employ several methods to ensure that our prompts are informative at each step. We begin with a system prompt which precedes all prompting with every model regardless of what module is being generated. This seeks to reduce output randomness by setting constraints for the LLMs to abide by. Our system prompt was structured as:
\begin{tcolorbox}[width=1.0\linewidth, halign=left, colframe=black, colback=white, boxsep=0.01mm, arc=1.5mm, left=2mm, right=2mm, boxrule=0.5pt]\footnotesize
Our goal is to provide complete Verilog modules based on user-provided specifications. Only the specified module is necessary, no testbenches or supplementary modules are needed. Examples of compiler and simulation errors for a given module may be provided, in which case we will proceed to correct the module.
All modules will be provided in their complete and correct form.
\end{tcolorbox}

\begin{figure}[t!]
    \centering
    \begin{subfigure}[b]{0.98\columnwidth}
    \begin{lstlisting}[]
include @prev_submods.v@   #contains previously generated Verilog submodules
@top_module@ = +"64-to-1 multiplexer"+
@prev_modules@ = [+"2-to-1 multiplexer"+,+"4-to-1 multiplexer"+]
@next_module@ = +"8-to-1 multiplexer"+
@next_io@ = +"mux8_1(input [2:0] sel, input [7:0] in, output reg out)"+
@prompt@ = *"We will be designing a* @<top_module>@ *in Verilog using hierarchical submodules. We have generated the following submodules:* @<prev_modules>@ *implemented as:*
@<prev_submods.v>@
*Please use the previous submodules to hierarchically generate a* @<next_module>@ *defined as:
module* @<next_io>@*;
//Insert code here
endmodule"*

@next_output@ = Feedback_Loop(prompt)
append @next_output@ to @prev_submods.v@
Move to next step in hierarchy
\end{lstlisting}
\vspace{-3mm}
    \caption{Pseudocode for automatic creation of a hierarchical prompt from a template which references previously generated hierarchy steps in the form of Verilog modules, as well as a call to the generative loop which creates the output module.}
    \label{fig:hier-code}
    \end{subfigure}
    \begin{subfigure}[b]{0.98\columnwidth}
    \begin{lstlisting}[language=Verilog]
module mux2_1(input in1, input in2, input select, output out);
    assign out = select ? in2 : in1;
endmodule
module mux4_1(input [1:0] sel, input [3:0] in, output out);
    wire out1, out2;
    // First level of multiplexer
    mux2_1 m1(.in1(in[0]), .in2(in[1]), .select(sel[0]), .out(out1));
    mux2_1 m2(.in1(in[2]), .in2(in[3]), .select(sel[0]), .out(out2));
    // Second level of multiplexer
    mux2_1 m3(.in1(out1), .in2(out2), .select(sel[1]), .out(out));
endmodule
\end{lstlisting}
\vspace{-3mm}
    \caption{The LLM-generated hierarchical Verilog module, \emph{prev\_submods.v}, which the above pseudocode includes as part of the prompt. Note that it contains all prior hierarchical modules, i.e., \emph{mux2\_1} and \emph{mux4\_1}.}
    \label{fig:hier-mod}
\end{subfigure}
\caption{Structure of a hierarchical step, which uses automated prompting and existing hierarchy to generate a new module.}
\label{fig:hier-example}
\end{figure}
We then use a benchmark-specific global prompt describing the overall design objective, akin to non-hierarchical approaches and, in the HDHP mode, we iteratively append a submodule prompt, one for each submodule in the list provided by the designer. Therefore, the first prompt provided to the LLM for a 64-to-1 multiplexer design looks as below, where the top-level prompt is in \textcolor{blue}{blue} and the first 
submodule prompt is in \textcolor{green}{green}. Note that for each submodule, we only provide its name and the submodule interface to the LLM. 

\begin{tcolorbox}[width=1.0\linewidth, halign=left, colframe=black, colback=white, boxsep=0.01mm, arc=1.5mm, left=2mm, right=2mm, boxrule=0.5pt]\footnotesize
\textcolor{blue}{We will be designing a 64-to-1 multiplexer in Verilog using hierarchical submodules.} \textcolor{green}{We begin by generating a 2-to-1 multiplexer with the following structure: module mux2\_1(in1, in2, select, out)}
\end{tcolorbox}



An example of automatic hierarchical prompt creation can be seen in Figure \ref{fig:hier-example}, with Figure \ref{fig:hier-code} showing the pseudocode instantiating the prompt template and Figure \ref{fig:hier-mod} representing the file containing the modules generated so far.

A subtlety in prompting is the distinction between conversational and non-conversational (or text completion) LLMs. 
In conversational LLMs, prior prompts and responses are automatically added to the LLM's context. 
For conversational LLMs, our hierarchical prompting approach starts with the global prompt that describes the top-level module and the first submodule. In subsequent iterations, we can simply request the next submodule since previously generated submodules are implicitly remembered in the LLM's context.


For text completion LLMs, however, we have to strategically insert responses 
from prior steps. 
We begin with the same global prompt as above that describes the top-level design and 
asks for the first submodule. For models with large context windows, we could simply include the code of all the previous submodules within the prompt as we progress. For many open-source LLMs with shorter context windows, however, we must implement what we call the \emph{relay prompt} technique as we progress deeper into the hierarchy,  as shown in Figure~\ref{fig:textcomp}.

 Specifically, for a given step, we include the full code of the submodule generated immediately prior to the current step, but only provide the module instantiation line for earlier steps. This becomes a list of all prior submodules that the LLM has access to, always including the complete elaboration of the most recently generated submodule. This allows ``leap-frogging" from less complex modules to more complex ones without running into length limits. Along with this prior context, we ask for the next submodule in the design hierarchy. For example, the final text-completion prompt for the decoder hierarchy can be seen in Figure \ref{fig:textcompfinal}. The full code for the 3-to-8 decoder is included, in addition to the module instantiation for the 2-to-4 decoder. This is sufficient information to improve performance. Though we give the example of the decoder for simplicity, relay prompting is an efficiency tool which becomes much more important for large hierarchies, like the 128-bit AES cipher which consists of hundreds of lines of code.
Most open-source LLMs are text-completion models, and this prompting method allows them to leverage hierarchy the same way a conversational LLM like GPT-4 would. All of our benchmarks use relay prompts by default for standardization and efficiency, though it is possible to include full prior submodule information at each step for smaller-scale hierarchies.


\begin{figure}[htbp]
\centering
\begin{subfigure}[b]{0.95\columnwidth}
\begin{lstlisting}[stringstyle = \color{blue},keywordstyle=\color{black}]
We will be designing a 5-to-32 decoder in Verilog using hierarchical submodules. We begin by generating a 2-to-4 decoder:
module decoder2to4(<human-sourced I/O>)
@<LLM-generated module>
endmodule@
We can then use that module hierarchically to generate a 3-to-8 decoder:
module decoder3to8(
\end{lstlisting}
\vspace{-3mm}
\caption{Initial prompt for 2-to-4 decoder and follow-up prompt for 3-to-8 decoder.}
\label{fig:textcompinit}
\end{subfigure}
\begin{subfigure}[b]{0.98\columnwidth}
\begin{lstlisting}[stringstyle = \color{blue},keywordstyle=\color{black}]
The following Verilog implements a 5-to-35 decoder utilizing hierarchical submodules. We have the following module(s) already available for use:
module decoder2to4(<human-sourced I/O>)
We can then use that module hierarchically to generate a 3-to-8 decoder:
module decoder3to8(@<LLM-generated I/O>)
<LLM-generated module>
endmodule@
We can then use these modules to hierarchically generate a 5-to-32 decoder:
module decoder5to32(
\end{lstlisting}
\caption{Final prompt for 5-to-32 decoder with compressed version of 2-to-4 decoder.}
\label{fig:textcompfinal}
\end{subfigure}
\caption{Example of Hierarchical Verilog Decoder Implementation using text-completion LLM. Model outputs are in \textcolor{blue}{blue}.}
\label{fig:textcomp}
\end{figure}

\section{Methods and Model Selection}
We selected eight relevant LLMs that are at the cutting edge of the field. Table \ref{tab:benchmark} shows the full list of models selected and their results on each of the module benchmarks. Of those eight, six are open-source. For those open-source models, all benchmarking inference was run on a single NVIDIA A100 80GB GPU. Inference for the GPT models was run using the OpenAI API. 
We lock two important model parameters during inference: temperature, which is commonly thought of as a ``creativity" value and determines how random the LLMs token generation is, and top-p, which sets a probability threshold for generated tokens, allowing only those tokens above the threshold to be selected from the probability distribution. Higher temperature and lower top-p lead to more random outputs, and vice-versa. Temperature values for each model were locked at 0.5, and top-p for each model was set at 0.9.  

We chose to test the unspecialized Llama 2~\cite{touvron2023llama2}, and the more recent Llama 3~\cite{llama3modelcard} models to see how generalist open-source LLMs compare to  specialized models. We hypothesized that even these models would see considerable performance improvements via hierarchical prompting. As a more specialized option, we include the Code Llama~\cite{roziere2023code} model which is fine-tuned on code. It is effective at HDL generation despite not being its intended purpose. We test a pair of LLMs fine-tuned for Verilog generation, namely VeriGen 16b~\cite{thakur2023verigen} 
and RTL-Coder~\cite{liu2024rtlcoder}.
We elected not to use the baseline models for each of these, as generalist Llama models of similar size are included in our list. Finally, we fine-tuned our own Code Llama-Verilog model to make the already competitive baseline Code Llama more effective. Our open-source contenders were compared against GPT-3.5 Turbo, GPT-4 black-box LLMs~\cite{openai_gpt-4_2023-1}.

\section{Results and Evaluation}
We ran each benchmark using the selected LLMs for 10 iterations per model-method-module combination to get a more statistical sense of how each model performs. The NH experiments still utilized the tool feedback loop for error fixing, but had no hierarchy applied to prompting. Ten iterations per module were allowed. We tracked both the \emph{pass@k} values and the wall-clock time it took to generate the outputs. The \emph{pass@k} metric is defined as the likelihood that one or more of the top-\emph{k} LLM-generated modules will pass the testbench ~\cite{chen2021evaluating}. Mathematically: 

\begin{equation}
    pass@k = 1-\begin{bmatrix}\frac{{n-c \choose k}}{{n \choose k}}\end{bmatrix}
    \label{eqn:passatk}
\end{equation}

\emph{n} is the number of generation attempts, \emph{c} is the number of correct attempts that pass testing, and \emph{k} is success threshold. 


\begin{table*}[htbp]
\setlength{\tabcolsep}{2.5pt}
\resizebox{\linewidth}{!}{%
\begin{tabular}{l|l||l|l||l|l||l|l||l|l||l|l||l|l||l|l||l|l} 
\hline
Comp. Results & LLM Used & \multicolumn{2}{l||}{Llama 2} & \multicolumn{2}{l||}{Code Llama} & \multicolumn{2}{l||}{VeriGen} & \multicolumn{2}{l||}{CL-Verilog} & \multicolumn{2}{l||}{RTL-Coder} & \multicolumn{2}{l||}{Llama 3} & \multicolumn{2}{l||}{GPT-3.5} & \multicolumn{2}{l}{GPT-4} \\ 
\hline
 & {\cellcolor[rgb]{0.898,0.898,0.898}}Parameters & \multicolumn{2}{l||}{{\cellcolor[rgb]{0.898,0.898,0.898}}13b} & \multicolumn{2}{l||}{{\cellcolor[rgb]{0.898,0.898,0.898}}13b} & \multicolumn{2}{l||}{{\cellcolor[rgb]{0.898,0.898,0.898}}16b} & \multicolumn{2}{l||}{{\cellcolor[rgb]{0.898,0.898,0.898}}13b} & \multicolumn{2}{l||}{{\cellcolor[rgb]{0.898,0.898,0.898}}7b} & \multicolumn{2}{l||}{{\cellcolor[rgb]{0.898,0.898,0.898}}8b} & \multicolumn{2}{l||}{{\cellcolor[rgb]{0.898,0.898,0.898}}\textasciitilde{} 20b*} & \multicolumn{2}{l}{{\cellcolor[rgb]{0.898,0.898,0.898}}\textasciitilde{} 1.8t*} \\ 
\hline
 & Open/Closed & \multicolumn{2}{l||}{Open} & \multicolumn{2}{l||}{Open} & \multicolumn{2}{l||}{Open} & \multicolumn{2}{l||}{Open} & \multicolumn{2}{l||}{Open} & \multicolumn{2}{l||}{Open} & \multicolumn{2}{l||}{Closed} & \multicolumn{2}{l}{Closed} \\ 
\hline
Module benchmark & {\cellcolor[rgb]{0.898,0.898,0.898}}\begin{tabular}[c]{@{}>{\cellcolor[rgb]{0.898,0.898,0.898}}l@{}}Prompt method\\Results\end{tabular} & {\cellcolor[rgb]{0.898,0.898,0.898}}NH & {\cellcolor[rgb]{0.898,0.898,0.898}}H & {\cellcolor[rgb]{0.898,0.898,0.898}}NH & {\cellcolor[rgb]{0.898,0.898,0.898}}H & {\cellcolor[rgb]{0.898,0.898,0.898}}NH & {\cellcolor[rgb]{0.898,0.898,0.898}}H & {\cellcolor[rgb]{0.898,0.898,0.898}}NH & {\cellcolor[rgb]{0.898,0.898,0.898}}H & {\cellcolor[rgb]{0.898,0.898,0.898}}NH & {\cellcolor[rgb]{0.898,0.898,0.898}}H & {\cellcolor[rgb]{0.898,0.898,0.898}}NH & {\cellcolor[rgb]{0.898,0.898,0.898}}H & {\cellcolor[rgb]{0.898,0.898,0.898}}NH & {\cellcolor[rgb]{0.898,0.898,0.898}}H & {\cellcolor[rgb]{0.898,0.898,0.898}}NH & {\cellcolor[rgb]{0.898,0.898,0.898}}H \\ 
\hline
\multirow{2}{*}{64-to-1 Multiplexer} & \begin{tabular}[c]{@{}l@{}}pass@1:\\pass@5:\end{tabular} & \begin{tabular}[c]{@{}l@{}}0.0\\0.0\end{tabular} & \begin{tabular}[c]{@{}l@{}}0.4\\0.976\end{tabular} & \begin{tabular}[c]{@{}l@{}}0.0\\0.0\end{tabular} & \begin{tabular}[c]{@{}l@{}}0.7\\1.0\end{tabular} & \begin{tabular}[c]{@{}l@{}}0.0\\0.0\end{tabular} & \begin{tabular}[c]{@{}l@{}}0.8\\1.0\end{tabular} & {\cellcolor[rgb]{0.678,0.835,1}}\begin{tabular}[c]{@{}>{\cellcolor[rgb]{0.678,0.835,1}}l@{}}0.3\\0.916\end{tabular} & \begin{tabular}[c]{@{}l@{}}0.8\\1.0\end{tabular} & \begin{tabular}[c]{@{}l@{}}0.0\\0.0\end{tabular} & \begin{tabular}[c]{@{}l@{}}0.4\\0.976\end{tabular} & \begin{tabular}[c]{@{}l@{}}0.1\\0.5\end{tabular} & {\cellcolor[rgb]{0.612,0.804,0.612}}\begin{tabular}[c]{@{}>{\cellcolor[rgb]{0.612,0.804,0.612}}l@{}}0.8\\1.0\end{tabular} & \begin{tabular}[c]{@{}l@{}}0.0\\0.0\end{tabular} & \begin{tabular}[c]{@{}l@{}}0.7\\1.0\end{tabular} & \begin{tabular}[c]{@{}l@{}}0.2\\0.78\end{tabular} & \begin{tabular}[c]{@{}l@{}}0.9\\1.0\end{tabular} \\ 
\hhline{~-||--||--||--||--||--||--||--||--}
 & {\cellcolor[rgb]{0.898,0.898,0.898}}Avg. Time (s): & {\cellcolor[rgb]{0.898,0.898,0.898}}621.54 & {\cellcolor[rgb]{0.898,0.898,0.898}}342.63 & {\cellcolor[rgb]{0.898,0.898,0.898}}634.01 & {\cellcolor[rgb]{0.898,0.898,0.898}}302.65 & {\cellcolor[rgb]{0.898,0.898,0.898}}642.11 & {\cellcolor[rgb]{0.898,0.898,0.898}}310.64 & {\cellcolor[rgb]{0.898,0.898,0.898}}573.21 & {\cellcolor[rgb]{0.898,0.898,0.898}}315.28 & {\cellcolor[rgb]{0.898,0.898,0.898}}407.88 & {\cellcolor[rgb]{0.898,0.898,0.898}}345.72 & {\cellcolor[rgb]{0.898,0.898,0.898}}521.04 & {\cellcolor[rgb]{0.898,0.898,0.898}}306.27 & {\cellcolor[rgb]{0.898,0.898,0.898}}606.43 & {\cellcolor[rgb]{0.898,0.898,0.898}}327.34 & {\cellcolor[rgb]{0.898,0.898,0.898}}1325.69 & {\cellcolor[rgb]{0.898,0.898,0.898}}507.54 \\ 
\hline
\multirow{2}{*}{5-to-32 Decoder} & \begin{tabular}[c]{@{}l@{}}pass@1:\\pass@5:\end{tabular} & \begin{tabular}[c]{@{}l@{}}0.0\\0.0\end{tabular} & \begin{tabular}[c]{@{}l@{}}0.5\\0.996\end{tabular} & \begin{tabular}[c]{@{}l@{}}0.0\\0.0\end{tabular} & \begin{tabular}[c]{@{}l@{}}0.8\\1.0\end{tabular} & \begin{tabular}[c]{@{}l@{}}0.1\\0.5\end{tabular} & \begin{tabular}[c]{@{}l@{}}0.7\\1.0\end{tabular} & \begin{tabular}[c]{@{}l@{}}0.1\\0.5\end{tabular} & {\cellcolor[rgb]{0.612,0.804,0.612}}\begin{tabular}[c]{@{}>{\cellcolor[rgb]{0.612,0.804,0.612}}l@{}}0.9\\1.0\end{tabular} & {\cellcolor[rgb]{0.678,0.835,1}}\begin{tabular}[c]{@{}>{\cellcolor[rgb]{0.678,0.835,1}}l@{}}0.2\\0.78\end{tabular} & \begin{tabular}[c]{@{}l@{}}0.7\\1.0\end{tabular} & \begin{tabular}[c]{@{}l@{}}0.0\\0.0\end{tabular} & \begin{tabular}[c]{@{}l@{}}0.7\\1.0\end{tabular} & \begin{tabular}[c]{@{}l@{}}0.0\\0.0\end{tabular} & \begin{tabular}[c]{@{}l@{}}0.8\\1.0\end{tabular} & \begin{tabular}[c]{@{}l@{}}0.4\\0.976\end{tabular} & \begin{tabular}[c]{@{}l@{}}1.0\\1.0\end{tabular} \\ 
\hhline{~-||--||--||--||--||--||--||--||--}
 & {\cellcolor[rgb]{0.898,0.898,0.898}}Avg. Time (s): & {\cellcolor[rgb]{0.898,0.898,0.898}}567.85 & {\cellcolor[rgb]{0.898,0.898,0.898}}274.33 & {\cellcolor[rgb]{0.898,0.898,0.898}}543.24 & {\cellcolor[rgb]{0.898,0.898,0.898}}310.41 & {\cellcolor[rgb]{0.898,0.898,0.898}}577.98 & {\cellcolor[rgb]{0.898,0.898,0.898}}333.83 & {\cellcolor[rgb]{0.898,0.898,0.898}}566.48 & {\cellcolor[rgb]{0.898,0.898,0.898}}329.65 & {\cellcolor[rgb]{0.898,0.898,0.898}}475.2 & {\cellcolor[rgb]{0.898,0.898,0.898}}361.73 & {\cellcolor[rgb]{0.898,0.898,0.898}}488.92 & {\cellcolor[rgb]{0.898,0.898,0.898}}301.97 & {\cellcolor[rgb]{0.898,0.898,0.898}}532.41 & {\cellcolor[rgb]{0.898,0.898,0.898}}215.33 & {\cellcolor[rgb]{0.898,0.898,0.898}}829.26 & {\cellcolor[rgb]{0.898,0.898,0.898}}379.65 \\ 
\hline
\multirow{2}{*}{32-bit Barrel Shifter} & \begin{tabular}[c]{@{}l@{}}pass@1:\\pass@5:\end{tabular} & \begin{tabular}[c]{@{}l@{}}0.0\\0.0\end{tabular} & \begin{tabular}[c]{@{}l@{}}0.0\\0.0\end{tabular} & \begin{tabular}[c]{@{}l@{}}0.0\\0.0\end{tabular} & \begin{tabular}[c]{@{}l@{}}0.2\\0.78\end{tabular} & \begin{tabular}[c]{@{}l@{}}0.0\\0.0\end{tabular} & \begin{tabular}[c]{@{}l@{}}0.3\\0.916\end{tabular} & \begin{tabular}[c]{@{}l@{}}0.0\\0.0\end{tabular} & {\cellcolor[rgb]{0.612,0.804,0.612}}\begin{tabular}[c]{@{}>{\cellcolor[rgb]{0.612,0.804,0.612}}l@{}}0.4\\0.976\end{tabular} & \begin{tabular}[c]{@{}l@{}}0.0\\0.0\end{tabular} & \begin{tabular}[c]{@{}l@{}}0.0\\0.0\end{tabular} & \begin{tabular}[c]{@{}l@{}}0.0\\0.0\end{tabular} & \begin{tabular}[c]{@{}l@{}}0.1\\0.5\end{tabular} & \begin{tabular}[c]{@{}l@{}}0.0\\0.0\end{tabular} & \begin{tabular}[c]{@{}l@{}}0.1\\0.5\end{tabular} & \begin{tabular}[c]{@{}l@{}}0.3\\0.917\end{tabular} & \begin{tabular}[c]{@{}l@{}}0.7\\1.0\end{tabular} \\ 
\hhline{~-||--||--||--||--||--||--||--||--}
 & {\cellcolor[rgb]{0.898,0.898,0.898}}Avg. Time (s): & {\cellcolor[rgb]{0.898,0.898,0.898}}322.47 & {\cellcolor[rgb]{0.898,0.898,0.898}}79.51 & {\cellcolor[rgb]{0.898,0.898,0.898}}301.68 & {\cellcolor[rgb]{0.898,0.898,0.898}}51.23 & {\cellcolor[rgb]{0.898,0.898,0.898}}401.65 & {\cellcolor[rgb]{0.898,0.898,0.898}}43.7 & {\cellcolor[rgb]{0.898,0.898,0.898}}296.54 & {\cellcolor[rgb]{0.898,0.898,0.898}}35.21 & {\cellcolor[rgb]{0.898,0.898,0.898}}256.32 & {\cellcolor[rgb]{0.898,0.898,0.898}}61.23 & {\cellcolor[rgb]{0.898,0.898,0.898}}309.22 & {\cellcolor[rgb]{0.898,0.898,0.898}}29.64 & {\cellcolor[rgb]{0.898,0.898,0.898}}312.08 & {\cellcolor[rgb]{0.898,0.898,0.898}}18.27 & {\cellcolor[rgb]{0.898,0.898,0.898}}450.66 & {\cellcolor[rgb]{0.898,0.898,0.898}}42.44 \\ 
\hline
\multirow{2}{*}{4x4 Systolic Array} & \begin{tabular}[c]{@{}l@{}}pass@1:\\pass@5:\end{tabular} & \begin{tabular}[c]{@{}l@{}}0.0\\0.0\end{tabular} & \begin{tabular}[c]{@{}l@{}}0.0\\0.0\end{tabular} & \begin{tabular}[c]{@{}l@{}}0.0\\0.0\end{tabular} & \begin{tabular}[c]{@{}l@{}}0.2\\0.78\end{tabular} & \begin{tabular}[c]{@{}l@{}}0.0\\0.0\end{tabular} & {\cellcolor[rgb]{0.612,0.804,0.612}}\begin{tabular}[c]{@{}>{\cellcolor[rgb]{0.612,0.804,0.612}}l@{}}0.5\\0.996\end{tabular} & \begin{tabular}[c]{@{}l@{}}0.0\\0.0\end{tabular} & \begin{tabular}[c]{@{}l@{}}0.5\\0.996\end{tabular} & \begin{tabular}[c]{@{}l@{}}0.0\\0.0\end{tabular} & \begin{tabular}[c]{@{}l@{}}0.1\\0.5\end{tabular} & \begin{tabular}[c]{@{}l@{}}0.0\\0.0\end{tabular} & \begin{tabular}[c]{@{}l@{}}0.0\\0.0\end{tabular} & \begin{tabular}[c]{@{}l@{}}0.0\\0.0\end{tabular} & \begin{tabular}[c]{@{}l@{}}0.0\\0.0\end{tabular} & \begin{tabular}[c]{@{}l@{}}0.0\\0.0\end{tabular} & \begin{tabular}[c]{@{}l@{}}0.5\\0.996\end{tabular} \\ 
\hhline{~-||--||--||--||--||--||--||--||--}
 & {\cellcolor[rgb]{0.898,0.898,0.898}}Avg. Time (s): & {\cellcolor[rgb]{0.898,0.898,0.898}}1358.99 & {\cellcolor[rgb]{0.898,0.898,0.898}}456.22 & {\cellcolor[rgb]{0.898,0.898,0.898}}1312.05 & {\cellcolor[rgb]{0.898,0.898,0.898}}422.45 & {\cellcolor[rgb]{0.898,0.898,0.898}}1472.91 & {\cellcolor[rgb]{0.898,0.898,0.898}}467.38 & {\cellcolor[rgb]{0.898,0.898,0.898}}1342.15 & {\cellcolor[rgb]{0.898,0.898,0.898}}481.27 & {\cellcolor[rgb]{0.898,0.898,0.898}}1021.44 & {\cellcolor[rgb]{0.898,0.898,0.898}}503.24 & {\cellcolor[rgb]{0.898,0.898,0.898}}1101.77 & {\cellcolor[rgb]{0.898,0.898,0.898}}325.18 & {\cellcolor[rgb]{0.898,0.898,0.898}}1276.39 & {\cellcolor[rgb]{0.898,0.898,0.898}}297.63 & {\cellcolor[rgb]{0.898,0.898,0.898}}2452.41 & {\cellcolor[rgb]{0.898,0.898,0.898}}402.26 \\ 
\hline
\multirow{2}{*}{UART 8-bit} & \begin{tabular}[c]{@{}l@{}}pass@1:\\pass@5:\end{tabular} & \begin{tabular}[c]{@{}l@{}}0.0\\0.0\end{tabular} & \begin{tabular}[c]{@{}l@{}}0.2\\0.78\end{tabular} & \begin{tabular}[c]{@{}l@{}}0.0\\0.0\end{tabular} & \begin{tabular}[c]{@{}l@{}}0.4\\0.976\end{tabular} & \begin{tabular}[c]{@{}l@{}}0.0\\0.0\end{tabular} & \begin{tabular}[c]{@{}l@{}}0.4\\0.976\end{tabular} & \begin{tabular}[c]{@{}l@{}}0.0\\0.0\end{tabular} & {\cellcolor[rgb]{0.612,0.804,0.612}}\begin{tabular}[c]{@{}>{\cellcolor[rgb]{0.612,0.804,0.612}}l@{}}0.6\\1.0\end{tabular} & \begin{tabular}[c]{@{}l@{}}0.0\\0.0\end{tabular} & \begin{tabular}[c]{@{}l@{}}0.2\\0.78\end{tabular} & \begin{tabular}[c]{@{}l@{}}0.0\\0.0\end{tabular} & \begin{tabular}[c]{@{}l@{}}0.4\\0.976\end{tabular} & \begin{tabular}[c]{@{}l@{}}0.0\\0.0\end{tabular} & \begin{tabular}[c]{@{}l@{}}0.7\\1.0\end{tabular} & \begin{tabular}[c]{@{}l@{}}0.0\\0.0\end{tabular} & \begin{tabular}[c]{@{}l@{}}0.8\\1.0\end{tabular} \\ 
\hhline{~-||--||--||--||--||--||--||--||--}
 & {\cellcolor[rgb]{0.898,0.898,0.898}}Avg. Time (s): & {\cellcolor[rgb]{0.898,0.898,0.898}}2482.17 & {\cellcolor[rgb]{0.898,0.898,0.898}}672.9 & {\cellcolor[rgb]{0.898,0.898,0.898}}2100.58 & {\cellcolor[rgb]{0.898,0.898,0.898}}614.07 & {\cellcolor[rgb]{0.898,0.898,0.898}}2603.12 & {\cellcolor[rgb]{0.898,0.898,0.898}}682 & {\cellcolor[rgb]{0.898,0.898,0.898}}2529.74 & {\cellcolor[rgb]{0.898,0.898,0.898}}673.21 & {\cellcolor[rgb]{0.898,0.898,0.898}}1763.06 & {\cellcolor[rgb]{0.898,0.898,0.898}}699.53 & {\cellcolor[rgb]{0.898,0.898,0.898}}1754.22 & {\cellcolor[rgb]{0.898,0.898,0.898}}573.64 & {\cellcolor[rgb]{0.898,0.898,0.898}}1800.45 & {\cellcolor[rgb]{0.898,0.898,0.898}}564.23 & {\cellcolor[rgb]{0.898,0.898,0.898}}3144.22 & {\cellcolor[rgb]{0.898,0.898,0.898}}752.76 \\ 
\hline
\multirow{2}{*}{AES Block Cipher} & \begin{tabular}[c]{@{}l@{}}pass@1:\\pass@5:\end{tabular} & \begin{tabular}[c]{@{}l@{}}0.0\\0.0\end{tabular} & \begin{tabular}[c]{@{}l@{}}0.0\\0.0\end{tabular} & \begin{tabular}[c]{@{}l@{}}0.0\\0.0\end{tabular} & \begin{tabular}[c]{@{}l@{}}0.2\\0.78\end{tabular} & \begin{tabular}[c]{@{}l@{}}0.0\\0.0\end{tabular} & \begin{tabular}[c]{@{}l@{}}0.3\\0.916\end{tabular} & \begin{tabular}[c]{@{}l@{}}0.0\\0.0\end{tabular} & {\cellcolor[rgb]{0.612,0.804,0.612}}\begin{tabular}[c]{@{}>{\cellcolor[rgb]{0.612,0.804,0.612}}l@{}}0.3\\0.916\end{tabular} & \begin{tabular}[c]{@{}l@{}}0.0\\0.0\end{tabular} & \begin{tabular}[c]{@{}l@{}}0.0\\0.0\end{tabular} & \begin{tabular}[c]{@{}l@{}}0.0\\0.0\end{tabular} & \begin{tabular}[c]{@{}l@{}}0.0\\0.0\end{tabular} & \begin{tabular}[c]{@{}l@{}}0.0\\0.0\end{tabular} & \begin{tabular}[c]{@{}l@{}}0.1\\0.5\end{tabular} & \begin{tabular}[c]{@{}l@{}}0.0\\0.0\end{tabular} & \begin{tabular}[c]{@{}l@{}}0.5\\0.996\end{tabular} \\ 
\hhline{~-||--||--||--||--||--||--||--||--}
 & {\cellcolor[rgb]{0.898,0.898,0.898}}Avg. Time (s): & {\cellcolor[rgb]{0.898,0.898,0.898}}3212 & {\cellcolor[rgb]{0.898,0.898,0.898}}783.54 & {\cellcolor[rgb]{0.898,0.898,0.898}}3201.78 & {\cellcolor[rgb]{0.898,0.898,0.898}}795.4 & {\cellcolor[rgb]{0.898,0.898,0.898}}3456.11 & {\cellcolor[rgb]{0.898,0.898,0.898}}809.54 & {\cellcolor[rgb]{0.898,0.898,0.898}}3203.55 & {\cellcolor[rgb]{0.898,0.898,0.898}}732.14 & {\cellcolor[rgb]{0.898,0.898,0.898}}2387.92 & {\cellcolor[rgb]{0.898,0.898,0.898}}960.29 & {\cellcolor[rgb]{0.898,0.898,0.898}}2603.51 & {\cellcolor[rgb]{0.898,0.898,0.898}}706.59 & {\cellcolor[rgb]{0.898,0.898,0.898}}2652.42 & {\cellcolor[rgb]{0.898,0.898,0.898}}694.71 & {\cellcolor[rgb]{0.898,0.898,0.898}}3822.64 & {\cellcolor[rgb]{0.898,0.898,0.898}}806.89 \\
\hline
\end{tabular}
}
\centering

\caption{Results on hierarchical design benchmarks for open-source and closed-source proprietary LLMs. The results for each model are separated by prompting method --- (i) flat, non-hierarchical (NH) or hierarchical (H) . We report \emph{pass@k} for \(n=10\) attempts, and the average generation time per benchmark. Top \textbf{open-source} hierarchical performers are in \textcolor{green}{green}, and top non-hierarchical performers are in \textcolor{blue}{blue}; generation time is the tie-breaker where needed.}
\label{tab:benchmark}
\end{table*}

Table \ref{tab:benchmark} shows the \emph{pass@k} values for each model-method-module permutation on the benchmark. Hierarchical prompting boosts performance on complex designs, 
and especially so on on weaker models that fail completely with standard NH prompting.
Consider the performance of Llama-2, which is not intended for code generation, much less HDL. Without hierarchical prompting, it fails at every task on the benchmark as it often cannot generate Verilog syntax. It fails on simpler submodules like 8-to-1 multiplexers. However, when guided hierarchically, it has the potential to succeed at even some complicated tasks. 

When applied to specialized models, the results are even more impressive. {Hierarchical prompting enables open-source LLMs to outperform standard flat prompting outputs from GPT-3.5 and GPT-4}. Furthermore,  these techniques on the GPT models yield further performance improvement, {allowing GPT-3.5 and GPT-4 them to succeed consistently on difficult modules}. Overall, every LLM sees improvement with hierarchical prompting.

\begin{figure}[htbp]
        \centering
        \includegraphics[width=9cm]{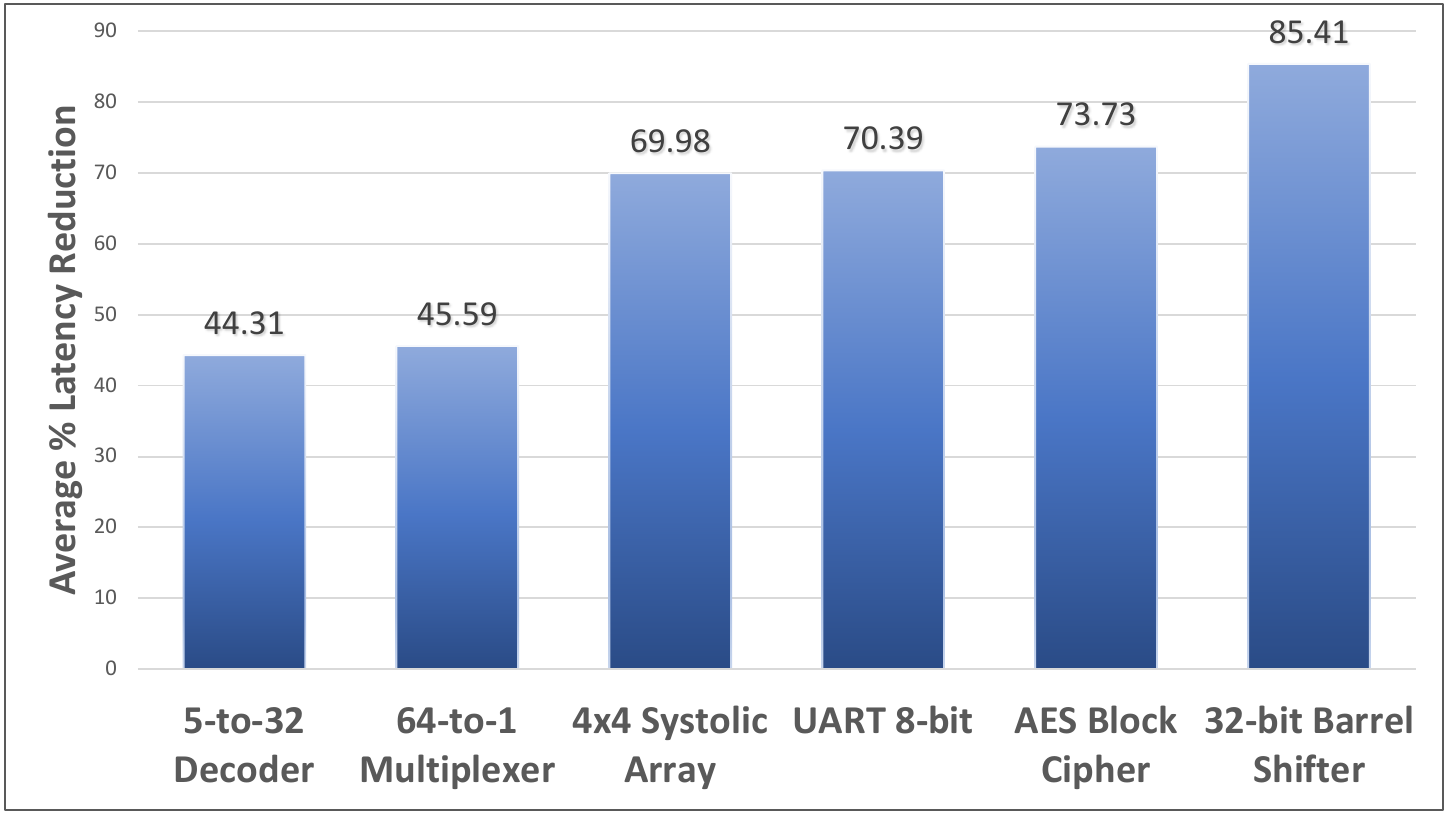}
        \caption{Hierarchical prompting yields consistent time savings vs. flat prompting, as seen by average \% latency reduction. More time savings are seen on modules which are difficult or impossible to generate non-hierarchically, or on those for which flat outputs are longer than hierarchical alternatives.}
        \label{fig:latency-reduction}
    \end{figure}

Hiearchical Prompting also reduces code generation time.
Figure \ref{fig:latency-reduction} reports the average percent time reduction due to hierarchical prompting across all LLMs for each of our benchmark. We see that the more difficult modules tend to have a much greater reduction in time, as successful hierarchical generation allows us to skip the lengthy error-handling process which sees the models re-generating past outputs and accounts for the majority of generation time.

\subsection{Purely Generative Results}
We implemented PGHP techniques for our benchmarks, but saw consistent failure for models \emph{except for} GPT-4. To diagnose the source, we selected 3 of the simplest benchmarks, and used a subset of our LLMs to generate 20 hierarchies each and evaluated them against the golden hierarchy plan from the HDHP version. Most LLMs performed inconsistently on this task, missing key submodules or inserting extraneous submodules (Table \ref{tab:pghp}).

On the other hand, {we find that GPT-4 significantly improves over flat prompting with PGHP}, as shown in Table \ref{tab:gpt4}.
PGHP is able to generate valid implementations for Systolic Array and UART on which GPT-4 fails completely in flat NH mode. Further, we see substantial gains in accuracy for the three simpler benchmarks. As we will see next, PGHP with GPT-4 is also successful in automatically designing a single-cycle
MIPS processor.


\begin{table}
\centering
\setlength{\extrarowheight}{0pt}
\addtolength{\extrarowheight}{\aboverulesep}
\addtolength{\extrarowheight}{\belowrulesep}
\setlength{\aboverulesep}{0pt}
\setlength{\belowrulesep}{0pt}
\resizebox{\linewidth}{!}{%
\begin{tabular}{|l||l||l||l||l||l||l|} 
\toprule
\begin{tabular}[c]{@{}l@{}}PGHP Accuracy\\by LLM\end{tabular} & Code Llama & VeriGen & CL-Verilog & RTL-Coder & GPT-3.5 & GPT-4 \\ 
\hline
\rowcolor[rgb]{0.902,0.902,0.902} Multiplexer & 0.15 & 0.15 & 0.25 & 0.05 & 0.2 & 0.85 \\ 
\hline
Decoder & 0.05 & 0.1 & 0.1 & 0.15 & 0.15 & 1.0 \\ 
\hline
\rowcolor[rgb]{0.902,0.902,0.902} Barrel Shifter & 0.05 & 0.0 & 0.1 & 0.0 & 0.0 & 0.35 \\
\bottomrule
\end{tabular}
}
\caption{Accuracy of LLM-decided architectural hierarchy out of 20 iterations. The inconsistency of most models when generating their own hierarchical plan is a major contributing factor to the failure of PGHP outside of GPT-4.}
\label{tab:pghp}
\end{table}

\begin{table}
\centering
\resizebox{\linewidth}{!}{%
\begin{tabular}{|l|l|l|l|l|l|l|} 
\hline
GPT-4 PGHP & Multiplexer & Decoder & Barrel Shift.  & Sys. Array & UART & AES \\ 
\hhline{|=======|}
pass@1 & 0.5 & 1.0 & 0.3 & 0.1 & 0.3 & 0.0 \\ 
\hline
pass@5 & 0.996 & 1.0 & 0.917 & 0.5 & 0.916 & 0.0 \\
\hline
\end{tabular}
}
\caption{\emph{pass@k} when applying purely generative hierarchical prompting (PGHP) to GPT-4 on each benchmark. Note improvement over flat prompting for most modules.}
\label{tab:gpt4}
\end{table}

\subsection{Identifying Common Failure Modes} 
We often see errors when LLMs generate text for too long and lose the original context of their goal. This occurs both for conversational and text-completion LLMs. We circumvented this by requesting no additional elements be generated via our system prompt, and re-inputting earlier context as a global prompt. Once a task is completed, text-completion LLMs tend to hallucinate. A common example is the unnecessary generation of testbenches or a random additional module. This is  avoided by truncating outputs at a useful end-token, usually the ``endmodule"  in Verilog.

Conversational LLMs can fall into ``perseverative" loops, a termed borrowed from neurology~\cite{BUCKINGHAM2008127},
continuing to repeat actions or words when the stimulus that brought on those behaviors has stopped, or when a competing stimulus has occurred that would normally trigger new behavioral routes.
One example is the continued use of an unnecessary always block when writing barrel shifter with non-hierarchical prompting, which can occur even when the LLM receives direct/detailed human feedback. As seen in Figure \ref{fig:persev} in the Appendix, the LLM will confirm it has done as asked, while continuing to output the same syntax as before. Avoiding such behavioral loops is another benefit of hierarchical prompting.

\section{Case Studies and Processor Generation}

To stress test our techniques, we hierarchically generated a full MIPS 16-bit single-cycle processor using GPT-3.5 and our Code Llama-Verilog model based on the PGHP paradigm. Flat prompting is unable to approach a functional processor design without considerable human oversight~\cite{blocklove2023chip}, but we hypothesized that hierarchy would bridge this gap and allow for automation. We tasked each model to first define a hierarchical structure for the processor as a list of submodules, then generate the processor stepwise. 

The models generated most necessary submodules, but missed key elements and struggled with assigning wire and signal names uniformly across modules, as seen in prior experiments. Tool feedback was helpful, but insufficient to bridge these issues. Ensuring all input and output wires/signals were named appropriately required human intervention and certain submodules like the control unit had to be directly requested, but all functional components were LLM-generated. After these interventions, we were able to synthesize the processors in Vivado, and successfully simulate processor instructions. We repeat this process by generating a RISC-V 32-bit processor utilizing GPT-4. Many of the issues present in GPT-3.5 are less problematic in GPT-4, and required much less human intervention. The newer model is better at wiring up interconnected modules and produces detailed descriptions of hierarchical architectures. 

To fully test this capability, we implemented the PGHP technique once more to generate another MIPS core via GPT-4 with no human intervention. After iterative tool feedback, GPT-4 converged on a synthesizeable processor that covered a version of the full MIPS ISA. Design and simulation results are shown in the Appendix. Figure \ref{fig:mips} shows the RTL and Figure \ref{fig:waveform} shows a waveform for this PGHP-sourced processor. 

We posit that this is the first-ever purely LLM-designed processor. That is, the design decisions were made entirely by the LLM with no human input, and all error handling was done automatically with tool feedback. Beyond the initial prompt of \emph{"Please define the necessary submodules in a 16-bit single cycle MIPS processor,"} no human design intervention was required.
We also see that the time taken to generate our processors is on the order of minutes, rather than multiple hours as seen in past methods. The time taken to complete the PGHP-based processor was 23 minutes, 37.85 seconds.

In order to get a sense of financial cost savings, we calculate the price-per-token and number of tokens generated when applying our hierarchical pipeline to GPT-3.5. We compare results for our full multiplexer hierarchy, our 32-bit barrel shifter, our MIPS processor, and our RISC-V processor. Table \ref{tab:financial} contains the full cost analysis. As one might expect, complex modules lead to higher costs and achieve more financial savings when generated hierarchically. 



\begin{table}
\centering
\resizebox{\linewidth}{!}{%
\begin{tabular}{|l|r|r|r|r|r|r|r|} 
\hline
 & \multicolumn{3}{c|}{Hierarchical} & \multicolumn{3}{c|}{Non-Hierarchical} & Savings \\
 & \multicolumn{1}{r}{I/P} & \multicolumn{1}{r}{O/P} & Cost & \multicolumn{1}{r}{I/P} & \multicolumn{1}{r}{O/P} & Cost & \% \\
 & \multicolumn{2}{c|}{(tokens)} & \$ & \multicolumn{2}{c|}{(tokens)} & \$ &  \\ 
\hline
\rowcolor[rgb]{0.937,0.937,0.937} Multiplexer & 92 & 2376 & 0.00484 & 91 & 3283 & 0.00666 & 27.23 \\ 
\hline
32-b Barrel & 262 & 1977 & 0.00422 & 191 & 4268 & 0.00873 & 51.69 \\ 
\hline
\rowcolor[rgb]{0.937,0.937,0.937} 16-b MIPS & 434 & 14226 & 0.02868 & 1243 & 31033 & 0.06314 & 54.58 \\ 
\hline
32-b RISC-V & 795 & 17310 & 0.03542 & 1593 & 42338 & 0.08593 & 58.8 \\
\hline
\end{tabular}
}
\caption{Cost of input (I/P) tokens processed and output (O/P) tokens generated of the modules using hierarchical and non-hierarchical prompting. Values based on GPT-3.5 tokenizer pricing.}
\label{tab:financial}
\end{table}

\section{Conclusion}
In this paper, we have proposed and evaluated Hierachical Prompting as a key tool for automated HDL code generation for complex modules. We show that with hierarchical prompting, even
smaller fine-tuned LLMs can correctly generate HDL for complex modules, when traditional flat prompting fails. 
On powerful models like GPT-4, hierarchical prompting is even more impressive, enabling the automatic generation of a single-cycle MIPS core. 
Overall, these methods give considerable insight into the potential of LLMs with either manually specified or automatically extracted design hierarchy.

There is considerable potential in this line of inquiry. We hope to include additional hardware design methods as part of a larger pipeline in the future. Considering the successes of methods like high-level synthesis (HLS), it stands to reason that leveraging different tools for different tasks could further improve results. We plan to fine-tune additional models with hierarchy in mind. Careful training dataset formulation could potentially lead to models which excel at hierarchical tasks, and may bridge the gap on PGHP performance for smaller models. We hope to expand evaluation resources for future benchmarking efforts to increase the strength of our \emph{pass@k} metric, ideally \(n=200\) samples per test.

\bibliographystyle{ACM-Reference-Format}
\bibliography{ref}
\clearpage
\onecolumn
\appendix
\section{Additional Figures}
\begin{table*}[htbp]
\centering
\begin{tabular}{|l|l|} 
\hline
Top Modules & Submodules \\ 
\hline
64-to-1 Multiplexer & \begin{tabular}[c]{@{}l@{}}2-to-1 mux\\4-to-1 mux\\8-to-1 mux\\16-to-1 mux\\32-to-1 mux\\64-to-1 mux\end{tabular} \\ 
\hline
5-to-32 Decoder & \begin{tabular}[c]{@{}l@{}}2-to-4 decoder\\3-to-8 decoder\\5-to-32 decoer\end{tabular} \\ 
\hline
32-bit Barrel Shifter & \begin{tabular}[c]{@{}l@{}}8-bit Barrel Shifter\\Rotation Control\\32-bit Barrel Shifter\end{tabular} \\ 
\hline
4x4 Systolic Array & \begin{tabular}[c]{@{}l@{}}Processor Element\\Control Logic\\Top-level 4x4 Array\end{tabular} \\ 
\hline
8-bit UART & \begin{tabular}[c]{@{}l@{}}Baud Rate Generator\\Receiver\\Transmitter\\State definitions\\Top-level UART\end{tabular} \\ 
\hline
128-bit AES Block Cipher & \begin{tabular}[c]{@{}l@{}}S-box\\Key Memory\\Encipher Block\\Decipher Block\\Control Logic\\Helper Functions(e.g.:\\inverse shift rows,\\mix columns, etc.)\end{tabular} \\
\hline
\end{tabular}
\caption{Top-level modules and corresponding submodules of the hierarchical benchmarks.}
\label{tab:bench_mods}
\end{table*}

\begin{figure}[htbp]
        \centering
    \includegraphics[width=\textwidth]{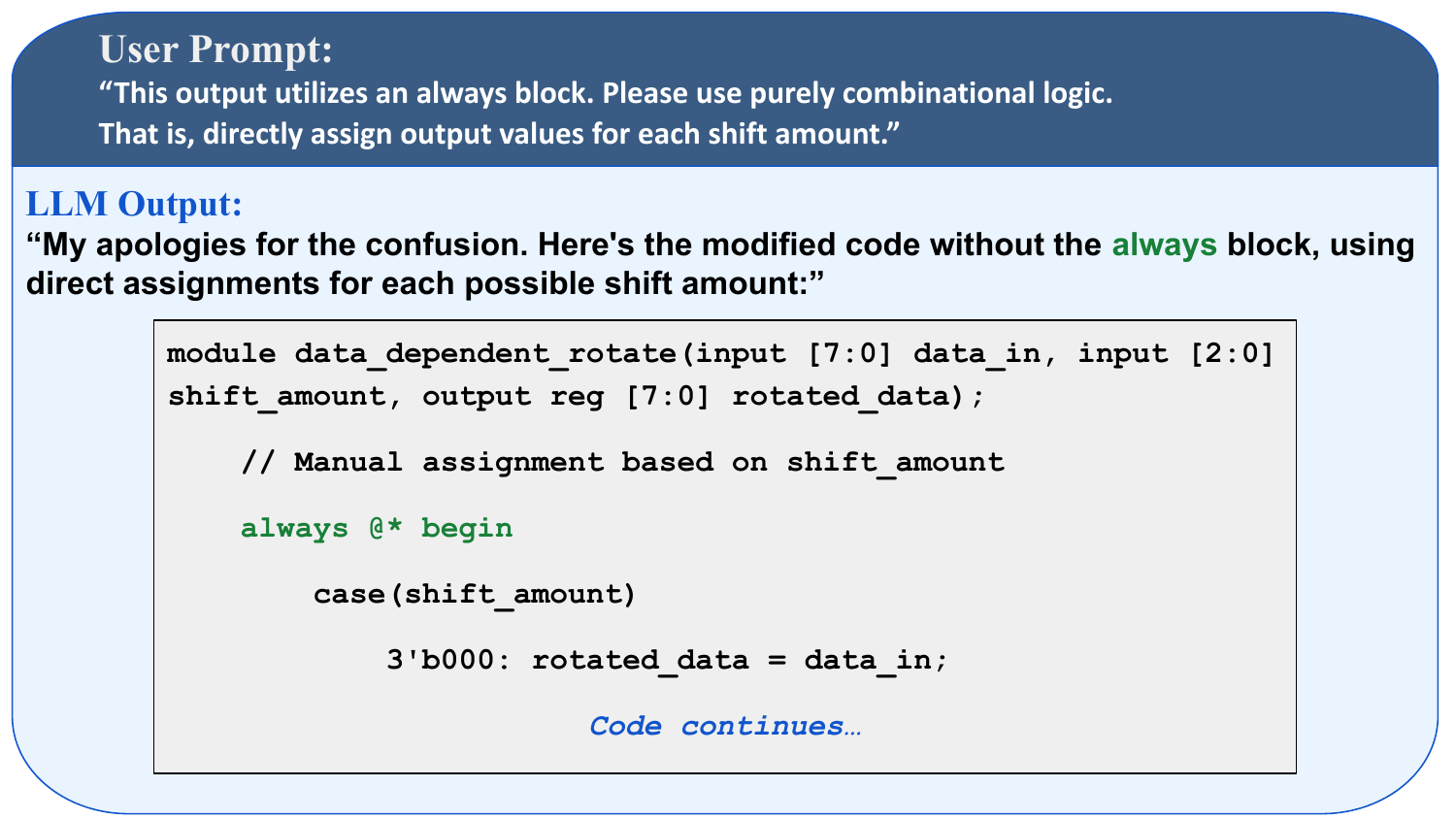}
        \caption{Perseveration-like behavior in GPT-3.5 when asked for rare syntax. }
        \label{fig:persev}
    \end{figure}

\begin{figure}[htbp]
        \centering
        \includegraphics[width=\textwidth]{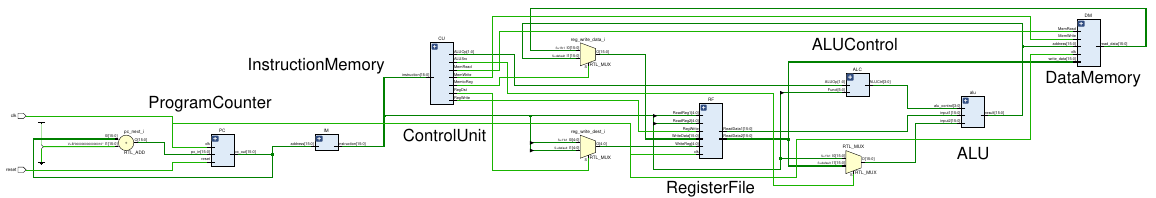}
        \caption{Elaborated Design Schematic of LLM-Generated MIPS Processor. Zoom in for submodule information.}
        \label{fig:mips}
        \end{figure}

        \begin{figure}[htbp]
        \includegraphics[width=\textwidth]{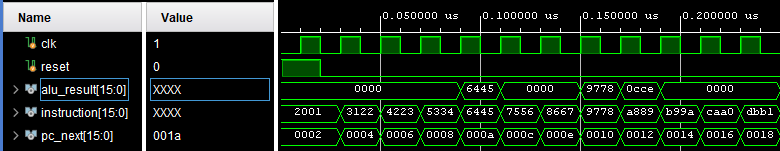}
        \caption{Sample waveform of running a series of instructions. Our first two instructions test the load and store functions, moving values between the registers and data memory. We then load two values from data memory, 12,006 and 13,663, into the registers R4 and R5 respectively. We then test the ADD function to sum them to 25669, or 6445 in hex. This intentionally echoes the value of the instruction for easy confirmation on the waveform. We then load two more values, 38,776 and 1,104 to test the OR instruction, to once again echo the instruction value of 9778 hex. We then test the SUBI instruction by subtracting 35,498 from 38,776 to get 3,278, or 0CCE hex. We then store all of our outputs thus far into data memory at various addresses to confirm successful saving. More thorough testing omitted for brevity. Though the LLM implements opcodes that are notably different from the standard MIPS ISA, all instructions are present and functional. Full instruction series: LW   R7, 8(R7); SW   R8, 9(R8); LW R4, 1(R4); LW R5, 3(R5); ADD R6, R4, R5; LW R7, 5(R7); LW R8, 6(R8); OR R9, R7, R8; SUBI R10, R9, 35498; SW R6, 9(R0); SW R9, 10(R0); SW R10, 11(R0);}
        \label{fig:waveform}
        \end{figure}
\section{Artifact Evaluation}
All relevant artifacts, including scripts, are publicly available at:

https://github.com/ajn313/ROME-LLM

They are additionally publicly archived at:

https://zenodo.org/records/13323449

https://figshare.com/projects/ROME-LLM/214771

We provide an example Python notebook to be used in Google Colab that includes all necessary elements for convenience. All that is
required is an OpenAI API key with GPT-4 usage enabled. We present a novel algorithm here in the form of our automated design pipeline.
It utilizes both an error handling feedback loop in addition to a hierarchical feedback loop in order to build on simpler module to make more
complex circuits \& structures. The benchmark we utilized is the novel Hierarchical ROME LLM benchmark described within the paper. This
includes a number of modules to be tested. We include unit tests to validate the performance of our tool.

All simulations are run using the open-source iVerilog tool which will be downloaded by the Colab notebook and will run via script calls
in the Colab environment. No other compiler/simulator is required. All software dependencies are handled by the Colab notebook, but a
list of dependencies will additionally be available on the GitHub. The notebook will default to using GPT-4 as the model, though this can
easily be changed to another OpenAI model. No download is necessary as it will run remotely through the OpenAI API. Instructions will be
included on how to instead use an open-source model, including our own CL-Verilog model, though it will be easiest to test using GPT.
Additionally, CL-Verilog will be released on Hugging Face.

Running all benchmarks may take numerous hours, but we include a subset of examples that should be less time consuming while
still displaying performance. Measurements can include pass@k, time to completion, or binary pass/fail. These values are affected by the
stochasticity of LLMs and may not match up perfectly to our own observations, for better or worse. Results should be reproducible via our
Colab notebook, though there is likely to be notable variance between runs, as our pass@k measurements show. The final output should be a
functional version of the target module which passes testbench simulation.

The testbenches can be downloaded from GitHub and uploaded to Colab, or cloned directly from the repo. No more than a few megabytes
of local disk space should be required if downloading is preferred, as the testbenches are on average around 2kb. One option for quickly
opening our notebook is githubtocolab.com, which will open the notebook directly in Google Colab from GitHub. Instructions are on the
tool’s site.

MIT license is used for code where applicable. All artifacts will be archived on Zenodo and FigShare and publicly available on GitHub at the above links.

Zenodo DOI:

https://doi.org/10.5281/zenodo.13323449
\end{document}